\documentclass[secnumarabic, graphics,floatfix, nofootinbib,tightenlines,nobibnotes, aps, prl, 12pt]{revtex4-2}
\usepackage[english]{babel}
\usepackage[utf8]{inputenc}
\usepackage{amsfonts}
\usepackage{multirow}
\usepackage{graphicx}
\usepackage{epstopdf}
\usepackage{hyperref}
\usepackage{latexsym}
\usepackage{amsmath}
\usepackage{amssymb}

\usepackage[utf8]{inputenc}
\hypersetup{
    colorlinks=true,
    linkcolor=red,
    citecolor=blue,
}
\usepackage{color}
\usepackage[T1]{fontenc}
\usepackage{txfonts}
\usepackage{mathrsfs}
\usepackage[center]{subfigure}

\begin{document}
 \setcounter{secnumdepth}{2}
 \newcommand{\bq}{\begin{equation}}
 \newcommand{\eq}{\end{equation}}
 \newcommand{\bqn}{\begin{eqnarray}}
 \newcommand{\eqn}{\end{eqnarray}}
 \newcommand{\nb}{\nonumber}
 \newcommand{\lb}{\label}
 
\title{High-order matrix method with delimited expansion domain}

\author{Kai Lin$^{1,2}$}\email[E-mail: ]{lk314159@hotmail.com}
\author{Wei-Liang Qian$^{2,3,4}$}\email[E-mail: ]{wlqian@usp.br}

\affiliation{$^{1}$ Hubei Subsurface Multi-scale Imaging Key Laboratory, Institute of Geophysics and Geomatics, China University of Geosciences, 430074, Wuhan, Hubei, China}
\affiliation{$^{2}$ Escola de Engenharia de Lorena, Universidade de S\~ao Paulo, 12602-810, Lorena, SP, Brazil}
\affiliation{$^{3}$ Faculdade de Engenharia de Guaratinguet\'a, Universidade Estadual Paulista, 12516-410, Guaratinguet\'a, SP, Brazil}
\affiliation{$^{4}$ Center for Gravitation and Cosmology, School of Physical Science and Technology, Yangzhou University, 225002, Yangzhou, Jiangsu, China}

\date{Feb. 25th, 2023}

\begin{abstract}
Motivated by the substantial instability of the fundamental and high-overtone quasinormal modes, recent developments regarding the notion of black hole pseudospectrum call for numerical results with unprecedented precision.
This work generalizes and improves the matrix method for black hole quasinormal modes to higher orders, specifically aiming at a class of perturbations to the metric featured by discontinuity intimately associated with the quasinormal mode structural instability.
The approach is based on the mock-Chebyshev grid, which guarantees its convergence in the degree of the interpolant.
In practice, solving for black hole quasinormal modes is a formidable task.
The presence of discontinuity poses a further difficulty so that many well-known approaches cannot be employed straightforwardly.
Compared with other viable methods, the modified matrix method is competent in speed and accuracy. 
Therefore, the method serves as a helpful gadget for relevant studies.

\end{abstract}

\maketitle

\newpage

\section{Introduction}\label{section1}

Owing to the valuable information extracted from the direct detection of the gravitational waves (GWs)~\cite{agr-LIGO-01, agr-LIGO-02, agr-LIGO-03, agr-LIGO-04}, the advent of GW astronomy is widely recognized as an inauguration of a novel era.
While the speculations to probe the strong-field regime of gravity have promoted rapid development in recent years, GW measurement is substantially affected by a few crucial factors.
Among others, frequency fluctuations continue to pose a primary challenge for the experimental implementations of space-borne laser interferometer projects~\cite{agr-LISA-01, agr-TianQin-01, agr-Taiji-01}.
In the literature, various noise suppression schemes were proposed~\cite{agr-TDI-review-01, agr-TDI-review-02, agr-TDI-Wang-01}, and the relevant detector sensitivity is formulated in terms of the signal-to-noise ratio~\cite{agr-SNR-Wang-01, agr-SNR-Wang-02}.
From the theoretical perspective, pertinent estimations have also been made regarding the viability of the black hole spectroscopy~\cite{agr-SNR-05, agr-SNR-06, agr-SNR-10, agr-SNR-18, agr-SNR-20, agr-SNR-36}.
Besides, as a possible source of gravitational radiations, a realistic black hole or neutron star is exposed to various surrounding matters, and subsequently, the spacetime deviates from that of an ideally symmetric metric.
As a result, the GWs might depart substantially from those emanating from an isolated compact object, giving rise to the notion of {\it dirty} black holes~\cite{agr-bh-thermodynamics-12, agr-qnm-33, agr-qnm-34, agr-qnm-54}.

In this context, significant efforts have been devoted to modeling systems composed of compact astrophysical objects, such as binaries of black holes or neutron stars.
Black hole quasinormal modes (QNMs)~\cite{agr-qnm-review-02, agr-qnm-review-03, agr-qnm-review-06} primarily constitute the ringdown stage of the merger process.
In terms of dissipative oscillations, these temporal profiles carry intrinsic properties of the underlying black hole spacetime, constraint by a few no-hair theorems~\cite{agr-bh-nohair-01, agr-bh-nohair-04}.
The scalar QNMs of dirty black holes was first investigated by Leung {\it et al.} for nonrotating metrics, where the deviations of quasinormal frequencies were evaluated by employing the generalized logarithmic perturbation theory.
A detailed analysis was later performed by Barausse {\it et al.} concerning a small perturbation about a central Schwarzschild black hole~\cite{agr-qnm-54}.
Regarding QNMs, the authors observed that the resultant QNMs might differ substantially from those of the isolated black hole.
It was concluded that the astrophysical environment would not significantly affect the black hole spectroscopy by using an appropriate template for the waveform.
Among the various scenarios explored in~\cite{agr-qnm-54}, the thin shell model was shown to give the most prominent modification to the QNM spectrum.

The above considerations are closely connected with the notion of black hole pseudospectrum, initiated and explored by Nollert~\cite{agr-qnm-35} and Nollert and Price~\cite{agr-qnm-36}.
It was demonstrated that the high-overtone modes of the QNM spectrum are significantly affected by a series of small-scale perturbations in terms of step functions. 
In other words, instability of the QNM spectrum against {\it ultraviolet} perturbations was observed, which undermines the understanding that once a reasonable approximation is adopted for the effective potential, the resulting QNMs will not deviate drastically.
Daghigh {\it et al.}~\cite{agr-qnm-50} confirmed the findings by considering a continuous piecewise approximation to the potential.
In~\cite{agr-qnm-lq-03}, we argued that as long as discontinuity is present, the asymptotical behavior of the QNM spectrum will be non-perturbatively modified.
Instead of climbing up the imaginary frequency axis~\cite{agr-qnm-continued-fraction-02, agr-qnm-continued-fraction-03}, the high-overtone modes will tend to stretch out along the real axis.
As was shown analytically, the result persists even when the discontinuity is located significantly further away from the horizon or arbitrarily insignificant. 
As a matter of fact, from the perspective that one views the quasinormal modes as nothing but the poles of the scattering matrix in the momentum space, the distribution of the poles is a renowned problem that has been explored by a few seminal works~\cite{qft-smatrix-03, qft-smatrix-04, qft-smatrix-05}.
In particular, the effect of discontinuity on the reflection coefficient was addressed by Berry~\cite{qft-smatrix-04}.
The matching condition for the waveform and its first-order derivative is essentially identical to Israel's junction condition employed in the present study (see Eq.~\eqref{Israel} below).
Jaramillo {\it et al.}~\cite{agr-qnm-instability-07, agr-qnm-instability-13, agr-qnm-instability-14} analyzed the problem by employing the notion of structural stability.
Specifically, the QNM spectrum's stability was explored in the context of randomized perturbations to the metric.
Using Chebyshev's spectral method in the hyperboloidal coordinates~\cite{agr-qnm-hyperboloidal-01}, it was found that the boundary of the pseudospectrum migrates toward the real frequency axis.
Reinforced by the existing results, such a conclusion indicates a universal instability of the high-overtone modes triggered by ultraviolet perturbations.
More recently, Cheung {\it et al.}~\cite{agr-qnm-instability-15} observed that even the fundamental mode can be destabilized under generic perturbations.
The results were further demonstrated by a simple toy model whose effective potential is also featured by discontinuity.

In the traditional black hole perturbation theory, one explores the stability of the underlying metric through the QNMs.
However, recent studies have revealed the significant implications of the QNM spectrum's stability since an insignificant change in the system might potentially cause a drastic modification to the QNM spectrum.
As a result, the QNMs extracted from an isolated black hole must be carefully scrutinized before they are used to form a template for the observational GW waveform.
On the practical side, these results call for numerical results with unprecedented precision.
Moreover, as pointed out in~\cite{agr-qnm-instability-07}, the relevant perturbations to the non-selfadjoint operator in question must be plausible so that the obtained results are physically meaningful.
To ascertain whether the system is beset with spectral instability, one needs to explore pertinent metric perturbations reminiscent of those occurring in a realistic environment that also warrants substantial impacts on the QNM spectrum.
Among other possibilities, a mathematically simple and physically relevant class of candidates are effective potentials that possess discontinuity.

Notably, many conventional approaches for the quasinormal modes cannot be straightforwardly applied to cases involving discontinuity.
For instance, the standard WKB formulae~\cite{agr-qnm-WKB-01, agr-qnm-WKB-02, agr-qnm-WKB-03, agr-qnm-WKB-04, agr-qnm-WKB-05, agr-qnm-WKB-06} evaluate the quasinormal frequencies using only the information of the effective potential in the vicinity of its maximum.
Also, the monodromy method~\cite{agr-qnm-40} resides on the assumption $\Im \omega \gg \Re \omega$, which becomes irrelevant for the present scenario one is dealing with QNMs with significant real parts.
The QNMs have been investigated for pulsating relativistic stars for which the discontinuity occurs at the surface.
In~\cite{agr-qnm-star-07}, Kokkotas and Schutz utilized the numerical integration, while Leins {\it el al.}~\cite{agr-qnm-star-08} modified Leaver's continued fraction method~\cite{agr-qnm-continued-fraction-01}. 
The latter turned out to be a dependable approach to handle discontinuity in the effective potential and has been adopted by a few authors in subsequential studies~\cite{agr-qnm-34, agr-qnm-star-25, agr-qnm-54}.
Moreover, an irregular singular point occurs at the horizon of a maximally charged Reissner-Nordstr\"om black hole.
As a result, the series expansion for the waveform that is employed by the continued fraction method becomes invalid.
In~\cite{agr-qnm-continued-fraction-09}, Onozawa {\it et al.} pointed out that the method is still applicable even to the extremal black holes if one expands the waveform about a suitable ordinary point.
As an improvement of Leaver's method, its applicability is not constrained by the specific type of singularity in the Regge-Wheeler potential. 
This contrasts Leaver's original approach, which cannot be applied when the equation possesses two irregular singular points at both boundaries.

The matrix method~\cite{agr-qnm-lq-matrix-01,agr-qnm-lq-matrix-02,agr-qnm-lq-matrix-03,agr-qnm-lq-matrix-04,agr-qnm-lq-matrix-05,agr-qnm-lq-matrix-06} is an approach that reformulates the QNM problem into a matrix equation for the complex frequencies.
The approach is reminiscent of the continued fraction method, and their main difference resides in the choice of the grid points where the expansion of the waveform is performed~\cite{agr-qnm-lq-matrix-01}.
It can also be viewed as a further generalization of the method proposed in~\cite{agr-qnm-continued-fraction-09}.
Besides the spherically symmetric cases~\cite{agr-qnm-lq-matrix-02}, the method can be straightforwardly generated to metrics with axial symmetry~\cite{agr-qnm-lq-matrix-03}.
In practice, the method is rather competent to deal with systems involving coupled degrees of freedom~\cite{agr-qnm-lq-matrix-07} or different sectors of the master equation are coupled~\cite{agr-qnm-lq-matrix-03}.
The method is also flexible to handle different boundary conditions~\cite{agr-qnm-lq-matrix-04}.
It has been generalized to deal with dynamic black hole spacetimes~\cite{agr-qnm-lq-matrix-05}.
More recently, the original method was generalized~\cite{agr-qnm-lq-matrix-06} to handle effective potentials containing discontinuity.
These features indicate that the method is a promising alternative in the toolkit for black hole QNMs.
The method matrix method is shown to offer reasonable accuracy as well as efficiency, and has been adopted in various studies~\cite{Destounis:2018utr, Destounis:2018qnb, Panotopoulos:2019tyg, Destounis:2019zgi, Hu:2019cml, Cardoso:2017soq, Liu:2019lon, agr-qnm-50, Shao:2020gwr, Lei:2021kqv, Zhang:2022roh, Li:2022khq, Shao:2022oqv, Mascher:2022pku}.
Nonetheless, it also entails some drawbacks.
Similar to the continued fraction method, the present scheme is sometimes constrained by the domain where the expansion is convergent.
Also, as one goes to a higher order, the results might be plagued by the Runge phenomenon, which must be handled with more care.
As the recent results regarding the instability of the fundamental mode~\cite{agr-qnm-instability-15} invite further studies regarding different metrics and perturbations, it is of increasing interest to explore the QNMs with much higher precision, which motivated us to push the matrix method to higher orders.
In this work, we investigate the precision of the matrix method at higher orders and analyze the convergence, precision, and efficiency of the method under such a circumstance.
In particular, a generalized high-order matrix method is implemented for black hole QNMs aiming at a specific class of metrics featured by discontinuity.
By comparing the results obtained by other approaches, we show that the modified matrix method competes with the task.

The remainder of the paper is organized as follows.
The following section gives an account of the matrix method and discusses its main features.
By comparing against other approaches, the precision and efficiency of the method at high order are analyzed in Sec.~\ref{section3}.
While the results are primarily satisfactory, we also elaborate on some of the potential issues of the algorithm.
To this end, in Sec.~\ref{section4}, the generalized matrix method is elaborated.
Based on the mock-Chebyshev grid, the approach benefits from both the convergence of the Chebyshev nodes and the facility of a uniform grid. 
In terms of numerical examples, convergence and improvement are demonstrated, specifically for the scenario of black hole effective potentials featured by discontinuity.
Further discussions and the concluding remarks are given in Sec.~\ref{section5}.

\section{The matrix method}\label{section2}

The main idea of the matrix method~\cite{agr-qnm-lq-matrix-01} is to express the wave function and its derivatives through interpolation in terms of the function values on the grid points so that an ordinary differential equation can be rewritten as a system of algebraic equations associated with the grid.
In the case of black hole perturbation theory, the master equation of the QNM problem typically possesses the form~\cite{agr-qnm-review-02}
\begin{equation}\lb{QNMaster}
\left[\frac{\partial^2}{\partial r_*^2}+\omega^2-V_{\mathrm{eff}}\right]\Psi = 0 ~,
\end{equation}
where $V_{\mathrm{eff}}$ is known as the effective potential, the spatial variable is mostly chosen as the tortoise coordinates $r_*$, a complex eigenvalue $\omega$ is to be determined, and the wave function $\Psi$ is subject to the physically appropriate boundary conditions.
The matrix method discretizes Eq.~\eqref{QNMaster} and transforms it to some non-standard matrix eigen equation~\cite{agr-qnm-lq-matrix-02,agr-qnm-lq-matrix-03}.
Taylor expansion is carried out for the wave function, largely reminiscent of Leaver's continued fraction approach~\cite{agr-qnm-continued-fraction-01}.
The primary difference resides in the fact~\cite{agr-qnm-lq-matrix-01} that, in place of an expansion around the horizon, a series of expansions are carried out at discrete grid points.
Besides, in principle, the grid does not necessarily feature an even distribution.
Such freedom can be appropriately adjusted, either to improve the {\it resolution} in the region of interest or to minimize the oscillations by taking the Chebyshev nodes~\cite{book-approximation-theory-Cheney-Light}.
In many practical cases, the roots of the resulting matrix equation can be obtained simultaneously by most non-linear equation solvers.

In what follows, we briefly review the discretization procedure.
For a univariate function $f(x)$ defined on a closed set $x\in [x_L, x_R]$, one carried out individual Taylor expansion at $N$ discrete grid points, namely, $x_1,x_2,\cdots,x_N$ with $x_1=x_L$ and $x_N=x_R$.
Without loss of generality, we perform an expansion about $x=x_2$ and then evaluate the function at the remaining grid points.
This gives rise to $N-1$ relations between function values and the derivatives at $x_2$:
\bqn
\lb{2}
\Delta{\cal F}=M D ,
\eqn
where
 \bqn
 \lb{3}
\Delta{\cal F}=\left(
    f(x_1)-f(x_2),
    f(x_3)-f(x_2),
    \cdots,
    f(x_j)-f(x_2),
    \cdots,
    f(x_{N})-f(x_2)
\right)^T ,
\eqn

\bqn
\lb{4}
M= \left(
  \begin{array}{cccccc}
    x_1-x_2 & \frac{(x_1-x_2)^2}{2} &\cdots & \frac{(x_1-x_2)^i}{i!} &\cdots & \frac{(x_1-x_2)^{N-1}}{{(N-1)}!} \\
    x_3-x_2 & \frac{(x_3-x_2)^2}{2} &\cdots & \frac{(x_3-x_2)^i}{i!} &\cdots & \frac{(x_3-x_2)^{N-1}}{{(N-1)}!} \\
        \cdots & \cdots & \cdots & \cdots & \cdots &\cdots \\
    x_j-x_2 & \frac{(x_j-x_2)^2}{2} &\cdots & \frac{(x_j-x_2)^i}{i!} &\cdots & \frac{(x_j-x_2)^{N-1}}{{(N-1)}!} \\
        \cdots & \cdots & \cdots & \cdots & \cdots &\cdots \\
    x_{N}-x_2 & \frac{(x_{N}-x_2)^2}{2} &\cdots & \frac{(x_{N}-x_2)^i}{i!} &\cdots & \frac{(x_{N}-x_2)^{N-1}}{{(N-1)}!} \\
  \end{array}
\right) ,
\eqn
and
\bqn
\lb{5}
D= \left(
    f'(x_2),
    f''(x_2),
    \cdots,
    f^{(k)}(x_2),
    \cdots,
    f^{({N})}(x_2)
\right)^T .
\eqn
When feasible, the above matrix equation can be reversed to rewrite the derivatives at $x_2$ in terms of the function values.
Using Cramer's rule, we have
\bqn
\lb{5a}
f'(x_2)= \det(M_1)/\det(M),\nb\\
f''(x_2)= \det(M_2)/\det(M) , \eqn
where $M_i$ is the matrix formed by replacing the $i$-th column of $M$ by the column matrix $\Delta{\cal F}$.
Furthermore, by permuting the $N$ points, we can rewrite all the derivatives at the above $N$ points as linear combinations of the function values at those points.
Substituting the derivatives into the ordinary differential equation in question, formally one obtains a homogeneous matrix equations with $f(x_1),\cdots,f(x_N)$ as its variables
\bqn
\lb{masterFormalMM}
{\cal G}{\cal F}=0 ~,
\eqn
where $\cal G$ is a $N \times N$ matrix and the column matrix $\cal F$ reads
 \bqn
 \lb{3}
{\cal F}=\left(
    f(x_1),
    f(x_2),
    f(x_3),
    \cdots,
    f(x_j),
    \cdots,
    f(x_{N})
\right)^T . 
\eqn

The matrix method can be adapted to various boundary conditions. 
First, one obtains the asymptotic form of the waveform by matching the ingoing and outgoing boundary conditions, respectively, at the horizon and outer spatial bound.
The master equation is then rewritten after appropriately subtracting such asymptotic forms from the original wavefunction~\cite{agr-qnm-lq-matrix-04}.
The above process is similar to the continued fraction method~\cite{agr-qnm-continued-fraction-01}.
However, in the case of the continued fraction method, the convergent criterion for the waveform is explicitly considered, expressed by the recurrence relations for the expansion coefficients.
Conversely, the expansion coefficients are truncated at a given order for the matrix method.
To proceed, one transforms the tortoise coordinate $r_*$ into $x$ whose domain is of finite range.
For instance, one may choose $x\in[0,1]$, where the boundaries are located at $x=0$ and $x=1$.
By introducing the above transforms, the master equation Eq.~\eqref{QNMaster} is turned into~\cite{agr-qnm-lq-matrix-04}
\bqn
\lb{masterMM}
H(\omega, x)R(x)=0 ~,
\eqn
with the boundary conditions
\bqn
\lb{4}
R(x=0)=C_0~~~\text{and}~~~R(x=1)=C_1 ~.
\eqn
As the asymptotic part of the wave function has already been subtracted, $C_0$ and $C_1$ are constants.
For convenience, we introduce
\bqn
\lb{5}
F(x)=R(x){x(1-x)} ~,
\eqn
and rewrites Eq.~\eqref{masterMM} into the form
\bqn
\lb{qnmeqf}
G(\omega, x)F(x)=0 ~,
\eqn
with more straightforward boundary conditions
\bqn
\lb{qnmbcf}
F(x=0)=F(x=1)=0 ~.
\eqn

We note that Eq.~\eqref{masterFormalMM} is nothing but the discretized version of the transformed master equation Eq.~\eqref{qnmeqf}.
A crucial step is to accommodate the boundary conditions Eq.~(\ref{qnmbcf}), which implies that
\bqn
\lb{8}
f(x_1)=f(x_N)=0~.
\eqn
This can be accommodated by replacing Eq.~\eqref{masterFormalMM} with
\bqn
\lb{qnmeqMatrix}
\overline{\cal G}\ {\cal F}=0~,
\eqn
where the matrix $\overline{\cal G}$ is defined by
\bq
\lb{10}
\overline{\cal G}_{i, j}= 
\left\{\begin{array}{cc}
\delta_{i, j},     &  i=1~\text{or}~N \cr\\
{\cal G}_{i, j}, &  i=2,3,\cdots,N-1
\end{array}\right. ~.
\eq
The matrix equation Eq.~\eqref{qnmeqMatrix} implies that the quasinormal frequencies $\omega$ satisfy
\bqn
\lb{qnmDet}
\det \overline{\cal G}(\omega) = 0~,
\eqn
where the corresponding eigenvector ${\cal F}$ furnishes the wave function $\Psi(r_*)$ of the original master equation.
In most cases, the roots of Eq.~(\ref{qnmDet}) can be obtained by the standard algebraic equation solver.
It is also noted that even though, in principle, Eq.~\eqref{qnmeqMatrix} might introduce additional irrelevant roots, it does not pose a serious problem, as long as they stay away from the low-lying quasinormal frequencies.

Furthermore, it was proposed~\cite{agr-qnm-lq-matrix-06} that the above algorithm can be adapted for the effective potential possessing discontinuity.
To proceed, one assigns the point of discontinuity to one of the grid points $x = x_c$.
In general relativity, a discontinuity in the radial coordinate corresponds to a discontinuity in a spherically symmetric surface.
This implies that the Taylor expansion is no longer valid in the entire domain.
When viewed as a limit of some physical scenario, it can be dealt with by Israel's junction condition~\cite{agr-collapse-thin-shell-03, book-general-relativity-Poisson}.
In other words, the wave functions on the two sides of discontinuity are related by~\cite{agr-qnm-34, agr-strong-lensing-shadow-35}
\bqn
\lb{Israel}
\lim_{\epsilon\to 0^+}\left[\frac{R'(x_c+\epsilon)}{R(x_c+\epsilon)}-\frac{R'(x_c-\epsilon)}{R(x_c-\epsilon)}\right]=\kappa \ ,
\eqn
and in particular, for the Schr\"odinger-type master equation Eq.~\eqref{QNMaster}, we have
\bqn
\lb{kappa}
\kappa = \lim_{\epsilon\to 0^+}\int_{x_c-\epsilon}^{x_c+\epsilon} V_{\mathrm{eff}}(x)dx \ .
\eqn
If one considers a moderate finite jump, the above relation simplifies to a vanishing Wronskian
\bqn
\lb{Wronskian}
W(\omega) \equiv R(x_c+\epsilon)R'(x_c-\epsilon) - R(z_x-\epsilon)R'(x_c+\epsilon) = 0~.
\eqn

Accordingly, one revises the matrix $\overline{\cal G}$ in Eq.~\eqref{10} to adequately take into account the above relations.
The Taylor expansion must only be performed for the intervals of $x$ where the potential is continuous.
This means that the matrix $\overline{\cal G}$ is {\it almost} broken into diagonal sections of block submatrices.
Also, the relation Eq.~\eqref{Israel} or~\eqref{Wronskian} will be implemented on a line shared by two relevant blocks.
For instance, if the boundary $x_c$ corresponds to the $i$th grid point, then both $i$th line and column $\overline{\cal G}$ will be occupied by both blocks.
To implement Israel's condition, one replaces the original line with the boundary condition, which eventually involves the values of the entire wave function on the grid points.
This is why the modified matrix $\overline{\cal G}$ is not entirely block diagonalized.
It is not difficult to understand because, otherwise, the resulting QNM spectrum would be constituted by a simple summation of those pertaining to individual blocks.

\section{Precision of the matrix method at higher order} \lb{section3}

In this section, we study the higher-order results of QNMs obtained by some of the well-known approaches. 
In particular, we explore the precision and efficiency of the matrix method in comparison with other methods.
For Schwarzschild black holes, the Regge-Wheeler potential reads
\begin{equation}\label{eqRW}
V_{\mathrm{eff}}^1=V_{\mathrm{RW}}(r)\equiv\left(1-\frac{r_h}{r}\right)\left[\frac{\ell(\ell+1)}{r^2}+(1-s^2)\frac{r_h}{r^3}\right] ,
\end{equation}
where $r$ is the radial coordinate, related to the tortoise coordinate by $r_*=\int\frac{dr}{1-\frac{r_h}{r}}$, $r_h$ is the location of the horizon, and $\ell$ corresponds to the angular momentum.
The variable $x$ is defined as $x=\frac{r-r_h}{r}$.
In the remainder of this paper, calculations are carried out for a few low-lying gravitational QNMs with $s=-2$, $r_h=1$, and different $\ell$.
We have set up a personal computer configured with an {\it Intel} {\it Xeon} W-2454 CPU @3.70GHz and $2\times 16.0$ GB {\it Hynix} 2666 DDR4 memory. 
The numerical algorithms are implemented using {\it Mathematica} 11.3 on {\it Windows} 10 pro 19044.1889, under which the computational time is estimated. 
The conclusion drawn from axial gravitational perturbations is also verified to be, by and large valid to other types of perturbations.
The obtained numerical results are compared with continued fraction method~\cite{agr-qnm-continued-fraction-01} up to $300$th order and the WKB method~\cite{agr-qnm-WKB-01} of 3rd~\cite{agr-qnm-WKB-02}, 6th~\cite{agr-qnm-WKB-03}, 9th, and 12th~\cite{agr-qnm-WKB-04, agr-qnm-WKB-05, agr-qnm-WKB-06} orders.
The results are presented in Tab.~\ref{tab1}-\ref{tab5}.

The results obtained by the WKB method for various orders are presented in Tab.~\ref{tab1}.
For the $9$th and $12$th order WKB methods, we utilize the public code released in Ref.~\cite{agr-qnm-WKB-06}.
Regarding the WKB method, the quasinormal frequencies are obtained by solving for specific roots of a nonlinear equation governed by the properties of the effective potential near its maximum.
As a result, the approach is very efficient. 
Even at higher order, the computation time of the WKB method is typically no more than a few seconds, much faster when compared with the continued fraction and matrix methods.
On the other hand, by comparing Tab.~\ref{tab1} with the first four rows of Tab.~\ref{tab2}-\ref{tab5}, some deviation is observed at the fourth significant figure.
Also, going to a higher order does not necessarily guarantee a more precise result, particularly for smaller $\ell$.

The results obtained using the continued fraction and matrix methods are presented in the first four rows of Tab.~\ref{tab2}-\ref{tab5}.
For the matrix method, convergence is manifestly shown as one goes to higher orders.
By comparing the two methods, the numerical results are broadly consistent.
At lower order, the results from the $100$th continued fraction agree with those from the matrix method adopting $N=25$ grid points, most of the time, by at least seven significant figures. 
As one goes to higher orders, the agreement is observed to improve.
The results from the $300$th continued fraction agree with those from the matrix method using $N=61$ grid points by more than twelve significant figures for the fundamental mode. 
In terms of computation time, the matrix method shows some advantages compared to the continued fraction method, which is already reasonably efficient.
Nonetheless, the computation time reported in Tab.~\ref{tab2}-\ref{tab3} does not include that to pre-evaluate the coefficients given by the r.h.s. of Eq.~\eqref{5a}, which typically becomes time-consuming at higher order. 

\begin{figure*}[htbp]
\centering
\includegraphics[scale=1.0]{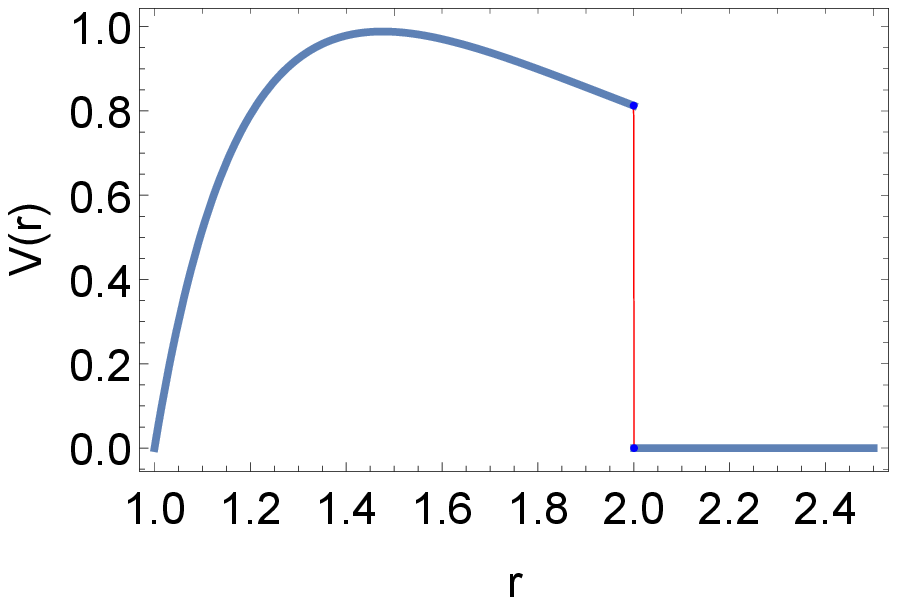} 
\caption{
The effective potential for axial gravitational perturbations in Schwarzschild black hole metric where a cut is implemented at $r_c=2$.}
\label{FigVcut}
\end{figure*}

However, when dealing with the effective potential of discontinuity, both methods encounter the problem of expansion convergence.
To put this into perspective, let us consider the effective potential $V_{\mathrm{eff}}$ that are obtained by truncating the Regge-Wheeler potential Eq.~\eqref{eqRW} at $r=r_c$.
As shown in Fig.~\ref{FigVcut}, one defines
\bqn
\lb{Veff2}
V_{\mathrm{eff}}^2=\left\{ \begin{matrix}
V_{\mathrm{RW}}& r\le r_c\\
0&r>r_c
\end{matrix} \right. \ ,
\eqn
For the {\it modified} continued fraction method~\cite{agr-qnm-star-08}, the series is known to become divergent if the expansion is carried out at $r_c < 2$.
The matrix method adopted for discontinuous effective potential will also lead to deviations at higher order.
To explicitly address this issue, we present, in the last four rows of Tab.~\ref{tab2}-\ref{tab3}, the calculated quasinormal frequencies by taking different values of $x_c=\frac{r_c-r_h}{r_c}>\frac12$ (so that $r_c > 2$).
The calculations are carried out using the recurrence Taylor expansion scheme~\cite{agr-qnm-lq-07}, an approach inspired by the modified continued fraction method, compared against those using the adopted matrix method~\cite{agr-qnm-lq-matrix-06}.
At higher orders $N\gtrsim 37$, the corresponding results obtained by the matrix method begin to present some inconsistent behavior.
This is understood that the matrix method is plagued by Runge's phenomenon~\cite{book-approximation-theory-Cheney-Light}, when applied to the discontinuous case.
From a mathematical perspective, the latter appears owing to the undesirable growth in the Lebesgue function.
Specifically, when using the Weierstrass approximation theorem to address the remainder in the Lagrange interpolation formula, the upper bound of the latter is governed by two factors: the cardinal function and the $(N+1)$th derivative of the waveform~\cite{book-numerical-analysis-Burden-Faires}.
In the case of a uniform grid, the Lebesgue constant $\Lambda_N$, which measures the bound of interpolation error, can be estimated according to Turetski~\cite{zmath-Runge-Chebyshev-02}.
By assuming specific grid numbers used in the above calculations, the rapid growth of the Lebesgue constant can be readily understood by taking the ratio $\Lambda_{37}/\Lambda_{25}\sim 2.5\times 10^3$.  

The above issues are also demonstrated by the numerical results regarding the convergence for different choices of grid points, presented in Tabs.~\ref{tab8}-\ref{tab10}.
As shown in the first six rows of Tab.~\ref{tab8}, the original matrix method~\cite{agr-qnm-lq-matrix-02} (denoted by MM0) is manifestly convergent as the grid number increases.
The results at higher order manifestly agree with the $300$th continued fraction by twelve significant figures for the fundamental mode. 
However, divergence occurs at higher order when the matrix method is generalized straightforwardly to deal with discontinuous effective potential~\cite{agr-qnm-lq-matrix-06}.
The problem appears for both the region $r_c\le 2$ and $2 < r_c <\infty$, as shown in the first six rows of Tabs.~\ref{tab9} and~\ref{tab10} (denoted by MM1).
Compared with those given in Tabs.~\ref{tab2} and~\ref{tab3}, the results are accurate for five or six significant figures when one is limited to moderate grid numbers.
As discussed above, Runge's phenomenon causes increasing deviations as the grid number increases.
To this end, in the following section, we proceed to discuss possible mitigations to the problems and present a generalized matrix method with a delimited expansion domain.

\section{Generalized matrix method and application to discontinuous effective potential}\lb{section4}

One of the well-known approaches to deal with the Runge phenomenon is to utilize the Chebyshev grid~\cite{book-approximation-theory-Cheney-Light}, for which the maximum of the nodal function is minimized.
Similarly, the Fekete grid maximizes the Vandermonde determinant, which in turn, also yields a smaller Lebesgue constant.
Moreover, various approaches based on equidistant nodes on different basis have also been proposed~\cite{zmath-Runge-Boyd-12}.
Nonetheless, many of these methods are featured by a specific nonlinear transform.
The latter enters into either the grid distribution or the regularization of the expansion coefficients.
Subsequently, the precision associated with the analytic form of Eq.~\eqref{5a} is undermined.
Besides, the spatial dependence of a differential equation owing to the Chebyshev discretization might become very {\it stiff}, which potentially leads to severe constraints.

In this section, we propose a generalized matrix method for black hole QNMs based on a uniform grid that primarily aims to suppress the Runge phenomenon.
The method is tailored for higher-order calculations with increasing precision while still warrants reasonable efficiency.
Such an approach follows the spirit of the mock-Chebyshev grid~\cite{zmath-Runge-mock-Chebyshev-01, zmath-Runge-mock-Chebyshev-02}.
The method was first introduced by Boyd and Xu, based on previous findings of Rakhmanov~\cite{zmath-Runge-Chebyshev-05}.
Mathematically, the Runge region, an area defined by error isosurface inside of which any pole of the waveform will lead to significant oscillations, shrinks as the ratio between the polynomial degree and grid points decreases.
It was shown that Chebyshev convergence could be achieved as long as the grid number grows at least as the square of the polynomial degree.
On the other hand, the above result can be understood intuitively regarding the Chebyshev grid: the clustering of the grid points near the end of the interpolation interval effectively suppresses the significant amplitude oscillations in the interpolant.
In particular, nodes' density is quadratic in the polynomial degree, in accordance with Rakhmanov's theorem~\cite{zmath-Runge-Chebyshev-05}.
Specifically, to interpolate a function by a polynomial of degree $N$, a mock-Chebyshev grid is defined as a subset of $N + 1$ points from an equispaced grid with $O(N^2)$ points chosen to mimic the non-uniform $N + 1$-point Chebyshev grid.
The authors of~\cite{zmath-Runge-mock-Chebyshev-01} made a more substantial claim that a good choice of expansion subset on an equispaced grid will guarantee a geometrical convergence in the degree of the interpolant.
Indeed, it was shown numerically that, for a moderate value of polynomial order, the error of mock-Chebyshev is mostly indistinguishable from the Chebyshev interpolant of the same degree.

Inspired by the above findings, the matrix method is adapted to this scenario by performing a delimited expansion on the grid.
To be specific, the function and its derivatives on the grid Eq.~\eqref{5a} are obtained using only a subset of grid $P$, which governs the polynomial order and satisfies~\cite{zmath-Runge-mock-Chebyshev-01}
\bqn
\lb{CriRes}
P< \sqrt{\frac1\chi} \sqrt{N}, 
\eqn
where
\bqn
\lb{chiLim}
\chi >\frac{2}{\pi^2} .
\eqn
Such a recipe can be readily applied to the case of effective potential with discontinuity discussed by the end of Sec.~\ref{section2}.

In what follows, we present the numerical results obtained using the generalized high-order matrix method.
We calculated both Regge-Wheeler effective potential Eq.~\eqref{eqRW} and that with discontinuity Eq.~\eqref{Veff2}.
The results are presented in the fifth to last rows of Tabs.~\ref{tab6} and~\ref{tab7}, and seventh to last rows of Tabs.~\ref{tab9} and~\ref{tab10} (denoted by MM2).
As shown in the fifth to last rows of Tabs.~\ref{tab6}, the proposed method agrees reasonably well with the recurrence Taylor expansion scheme at a relatively lower order.
The only deviation is observed when $x_c\sim 1$.
On the other hand, the computation time of the matrix method is comparable to or faster than the continued fraction method.
Tabs.~\ref{tab9} and~\ref{tab10}, we show the convergence of the approach as one goes to a higher order.
For the two truncation points considered there, desirable precision has been achieved.
The generalized high-order matrix method agrees with the recurrence Taylor expansion scheme by at least twelve significant figures.

Last but not least, in Tabs.~\ref{tab11} and~\ref{tab12}, we compare high-order results between the different methods elaborated in the present study.
The calculations are carried out using the WKB approximation, the continued fraction method, the original matrix method, and the generalized one at different orders.
The calculations are carried out for the first two low-lying modes with $\ell=2$, $8$, and $12$, regarding the Regge-Wheeler potential with $s=-2$ and $r_h=1$.
For the continuous effective potential in question, the continued fraction method, both versions of the matrix method demonstrate a reasonable degree of consistency at high orders.
As discussed, the high-order WKB method might become unstable for specific scenarios.
As expected, the performance is much improved at more significant angular momentum.
Nonetheless, minor discrepancies are still observed compared to continued fraction and matrix methods.

\section{Further discussions and concluding remarks} \label{section5}

To summarize, in this paper, we generalized the matrix method for black hole QNMs to higher order and to cope with discontinuous effective potential.
Mathematically, this is implemented by adopting the approach of a mock-Chebyshev grid.
The proposed approach achieves desirable precision by suppressing the Runge phenomenon and the analytic matrix coefficients by retaining a uniform grid.
Moreover, the numerical calculations' computation time is more favorable compared to its predecessor.
This is demonstrated by a detailed comparison of the obtained numerical results against other method.

In particular, when compared with the continued fraction method, widely recognized as the most precise method for black hole QNMs, the matrix method is shown to provide competent performance.
Moreover, unlike the continued fraction method, the matrix method does not require reformulating the master equation into a system of recurrence relation between the expansion coefficients.
This provides further flexibility when applying the matrix method to specific scenarios.

As Runge demonstrated more than a century ago, using evenly spaced grid points often leads to a severe convergence problem in the polynomial interpolation of a function. 
In this regard, the matrix method for black hole QNMs shows surprising robustness for continuous effective potential explored in the literature and this paper.
This might be understood as follows.
The master equation of black hole QMNs is of second order, and it is only discretized and utilized at the grid points.
Therefore, the relevant deviations, which also occur at those grid points, remain primarily restricted.
The problem becomes more significant in the presence of discontinuity, and the mock-Chebyshev is implemented in this regard to alleviate the divergence.

The main limitation of the current approach resides in the calculations of high-overtone modes.
Regarding the latter, the continued fraction method can precisely evaluate these modes when adapted to appropriate modifications~\cite{agr-qnm-continued-fraction-02, agr-qnm-continued-fraction-03}.
Besides, some dedicated approaches, such as Motl and Neitzke's monodromy method~\cite{agr-qnm-40} and high-overtone WKB approximation~\cite{agr-qnm-WKB-07, agr-qnm-WKB-08}, were developed aiming at these asymptotic states.
However, to our knowledge, neither of these methods can be straightforwardly employed to deal with discontinuity.
Although challenging, exploring the high overtone modes in metrics with a discontinuity is of physical interest.
We understand that the matrix method is a potential candidate for such a task.

Before closing this section, it is also worthwhile to briefly comment on a few recent developments in the QNM technique.
Machine learning method, particularly the artificial neural network, has been utilized to evaluate the QNMs.
Mathematically, the artificial neural network is an optimization scheme that can be adopted to solve eigenvalue problems~\cite {zml-neural-network-05}.
In Ref.~\cite{agr-qnm-60}, the method was adopted for black hole QNMs and exercised for four-dimensional pure dS and five-dimensional Schwarzschild AdS black holes.
Good agreement was manifestly obtained when compared with other methods.
For rotating black holes where the master equations typically consist of coupled equations associated with different degrees of freedom, a perturbative double expansion method was proposed~\cite{agr-qnm-61}.
The authors considered the second order in rotation and the first order in non-radial deviations, and the master equation describes the polar-led and axial-led perturbations.
The quasinormal frequencies are identified by the zeroes of the Wronskian, and good accuracy was achieved.
Apart from the technical challenge, these types of metrics are physically relevant owing to the specific feature known as eigenvalue repulsions~\cite{agr-qnm-63}.
A new method for QNM was proposed in~\cite{agr-qnm-62}.
The approach is based on the intriguing connection between gravity, gauge, and quantum integrable theories.
The authors pointed out that the resulting QNMs, incarnated in terms of a specific quantization condition, can be identified with the Bethe roots. 
Nonetheless, these advances are primarily associated with the waveforms in continuous background metrics.

On the other hand, the present study is primarily motivated by the recent development regarding structural instability in black hole QNMs which calls for high precision calculations, particularly those for effective potentials with discontinuity.
Apart from its mathematical simplification, discontinuity is a physically relevant scenario in black hole physics.
For example, the $w$-modes in pulsating neutron stars~\cite{agr-qnm-star-07, agr-qnm-star-08, agr-qnm-star-09, agr-qnm-star-27} are understood to be caused by the discontinuity in matter distribution, that serves as a concrete example of application.
For exotic compact objects, discontinuous matter distribution was introduced as one constructs the throat of traversable wormholes using the cut-and-paste procedure~\cite{agr-wormhole-10}.
Moreover, discontinuity plays an essential role from a dynamic perspective.
In the framework of $\Lambda$CDM model, {\it cusp} was found in the halo profile~\cite{agr-dark-matter-06, agr-dark-matter-07}, which are largely compatible with the presence of dark halos~\cite{agr-dark-matter-08}.
A discontinuous {\it splashback} feature takes place in the outer region of the halo, related to the sudden drop in the density profile~\cite{agr-dark-matter-21, agr-dark-matter-24}.
In the time evolution of a spherically collapsing matter, the interior metric was also featured by discontinuity~\cite{agr-collapse-thin-shell-11}.
Besides the QNM structural instability, discontinuity in the metric has been explored for other interesting implications. 
It was shown that discontinuity in the effective potential furnishes a possible origin of black hole echoes~\cite{agr-qnm-echoes-20}, associated with modifying the pole structure of Green's function.
A discontinuous dusty thin shell might affect the black hole shadow non-trivially~\cite{agr-strong-lensing-shadow-35}. 
Recent developments regarding the instability of the fundamental mode~\cite{agr-qnm-instability-15} invite further studies regarding different metrics and perturbations.
As the presence of discontinuity poses a difficulty for direct applications of most approaches for quasinormal modes, the proposed method serves as an alternative tool for relevant studies.

\section*{Acknowledgments}
We are grateful for the comments and suggestions from the two anonymous referees.
This work is supported by the National Natural Science Foundation of China (NNSFC) under contract Nos. 42230207, 11805166, 11925503, and 12175076.
We also gratefully acknowledge the financial support from
Funda\c{c}\~ao de Amparo \`a Pesquisa do Estado de S\~ao Paulo (FAPESP),
Funda\c{c}\~ao de Amparo \`a Pesquisa do Estado do Rio de Janeiro (FAPERJ),
Conselho Nacional de Desenvolvimento Cient\'{\i}fico e Tecnol\'ogico (CNPq),
Coordena\c{c}\~ao de Aperfei\c{c}oamento de Pessoal de N\'ivel Superior (CAPES),
A part of this work was developed under the project Institutos Nacionais de Ci\^{e}ncias e Tecnologia - F\'isica Nuclear e Aplica\c{c}\~{o}es (INCT/FNA) Proc. No. 464898/2014-5.
This research is also supported by the Center for Scientific Computing (NCC/GridUNESP) of S\~ao Paulo State University (UNESP).

\bibliographystyle{h-physrev}
\bibliography{references_qian, references_mm_citations}

\newpage

\begin{table*}[ht]
\caption{\label{tab1} Numerical results of $n$th low-lying QNMs for $s=-2$, $r_h=1$ and for different $\ell$.
The calculations are carried out using the $k$th order WKB method, denoted by WKB$_k$.}
\centering
\begin{tabular}{c c c c c}
         \hline\hline
& $\ell =2$, $n=0$ & $\ell =2$, $n=1$ & $\ell =3$, $n=0$ & $\ell =3$, $n=1$ \\
        \hline
$\text{WKB}_3$ &~~~$0.746324 - 0.178435i$~~~&~~~$0.692035 - 0.549831i$~~~&~~~$1.19853 - 0.185457i$~~~&~~~$1.16471 - 0.562812i$~~~\\
$\text{WKB}_6$ &$0.747239 - 0.177782i$&$0.692593 - 0.54696i$&$1.19889 - 0.185405i$&$1.16528 - 0.562581i$\\
$\text{WKB}_9$ &$0.747664 - 0.177341i$&$0.693528 - 0.543416i$&$1.19888 - 0.185375i$&$1.16507 - 0.562164i$\\
$\text{WKB}_{12}$ &$0.170309 + 79.993i$&$0.810392 + 420.933i$&$0.00505662 + 7.74397i$&$0.143984 + 38.8067i$\\
\hline
\hline
\end{tabular}
\end{table*}

\begin{table*}[ht]
\caption{\label{tab2} Computation time and numerical results of $n$th low-lying QNMs for $s=-2$, $r_h=1$ and for different $\ell$.
The calculations are carried out using the $100$th order continued fraction method.
The results for the original Regge-Wheeler effective potential Eq.~\eqref{eqRW} are presented in the first four rows denoted by $x_c=1$.
For the case where the truncation point $x_c \ne 1$, one considers the discontinuous effective potential defined in Eq.~\eqref{Veff2}.}
\centering
\begin{tabular}{c c c}
         \hline\hline
 & $\text{Time}$ & $\omega$ \\
\hline
\hline
$\ell=2$, $n=0$, $x_c=1$ ~~~&~~~$22.9018657s$~~~&~~~$0.74734338640252294366 - 0.17792462507976131814i$\\
$\ell=2$, $n=1$, $x_c=1$ ~~~&~~~$22.9565791s$~~~&~~~$0.69341608509169484572 - 0.54783348134071378818i$\\
$\ell=3$, $n=0$, $x_c=1$ ~~~&~~~$23.1517763s$~~~&~~~$1.19888657675535801472 - 0.18540609598072215335i$\\
$\ell=3$, $n=1$, $x_c=1$ ~~~&~~~$22.6802094s$~~~&~~~$1.16528764807123335457 - 0.56259622394483836164i$\\
$\ell=2$, $n=0$, $x_c=\frac78$ ~~~&~~~$27.3636992s$~~~&~~~$0.75375900567010947824 - 0.18133972924323434997i$\\
$\ell=2$, $n=0$, $x_c=\frac34$ ~~~&~~~$36.5881189s$~~~&~~~$0.79425305766566536385 - 0.14837000709837277317i$\\
$\ell=2$, $n=0$, $x_c=\frac58$ ~~~&~~~$31.6841693s$~~~&~~~$0.80929316810176245225 - 0.25418852805024597175i$\\
$\ell=2$, $n=0$, $x_c=\frac12$ ~~~&~~~$31.4066058s$~~~&~~~$0.72595071361390402716 - 0.35517954711880666267i$\\
$\ell=2$, $n=0$, $x_c=\frac{7}{16}$ ~~~&~~~$27.4044890s$~~~&~~~$0.65578027742727507282 - 0.39161380891220111843i$\\
$\ell=2$, $n=0$, $x_c=\frac{3}{8}$ ~~~&~~~$27.4154933s$~~~&~~~$0.57015231756261314027 - 0.41480639631613968928i$\\
$\ell=2$, $n=0$, $x_c=\frac{5}{16}$ ~~~&~~~$27.0177373s$~~~&~~~$0.47128814128866042242 - 0.42258134582724513639i$\\
$\ell=2$, $n=0$, $x_c=\frac{1}{4}$ ~~~&~~~$26.6673649s$~~~&~~~$0.36073050336278803859 - 0.41321763684510711044i$\\
\hline
\hline
\end{tabular}
\end{table*}

\begin{table*}[ht]
\caption{\label{tab3} The same as Tab.~\ref{tab2}, but the calculations are carried out using $300$th order continued fraction method.}
\centering
\begin{tabular}{c c c}
         \hline\hline
 & $\text{Time}$ & $\omega$ \\
\hline
\hline
$\ell=2$, $n=0$, $x_c=1$ ~~~&~~~$361.1824177s$~~~&~~~$0.74734336883598689863 - 0.17792463137781263197i$\\
$\ell=2$, $n=1$, $x_c=1$ ~~~&~~~$357.3262632s$~~~&~~~$0.69342199273744964125 - 0.54782975033113425721i$\\
$\ell=3$, $n=0$, $x_c=1$ ~~~&~~~$948.1775553s$~~~&~~~$1.19888657687498012275 - 0.18540609588989520951i$\\
$\ell=3$, $n=1$, $x_c=1$ ~~~&~~~$359.7117129s$~~~&~~~$1.16528760606654767305 - 0.56259622686999972572i$\\
$\ell=2$, $n=0$, $x_c=\frac78$ ~~~&~~~$1565.0179058s$~~~&~~~$0.78475423277016743212 - 0.13325759012490916480i$\\
$\ell=2$, $n=0$, $x_c=\frac34$ ~~~&~~~$720.4296342s$~~~&~~~$0.79425298413668122287 - 0.14836993024971837407i$\\
$\ell=2$, $n=0$, $x_c=\frac58$ ~~~&~~~$847.0548046s$~~~&~~~$0.80929316810176181597 - 0.25418852805024671575i$\\
$\ell=2$, $n=0$, $x_c=\frac12$ ~~~&~~~$836.4980019s$~~~&~~~$0.72595071361390402716 - 0.35517954711880666267i$\\
$\ell=2$, $n=0$, $x_c=\frac{7}{16}$ ~~~&~~~$721.6310936s$~~~&~~~$0.65578027742727507282 - 0.39161380891220111843i$\\
$\ell=2$, $n=0$, $x_c=\frac{3}{8}$ ~~~&~~~$838.6370523s$~~~&~~~$0.57015231756261288992 - 0.41480639631614803108i$\\
$\ell=2$, $n=0$, $x_c=\frac{5}{16}$ ~~~&~~~$711.7124568s$~~~&~~~$0.47128814128866042242 - 0.42258134582724513639i$\\
$\ell=2$, $n=0$, $x_c=\frac{1}{4}$ ~~~&~~~$708.5541854s$~~~&~~~$0.36073050336278803859 - 0.41321763684510711044i$\\
\hline
\hline
\end{tabular}
\end{table*}

\begin{table*}[ht]
\caption{\label{tab4} Computation time and numerical results of $n$th low-lying QNMs for $s=-2$, $r_h=1$ and for different $\ell$.
The calculations are carried out by the matrix method using $N=25$ grid points.
The computation time reported in this table does not include that to pre-evaluate the coefficients of Eq.~\eqref{5a}, which is typically time-consuming at higher order. 
The results for the original Regge-Wheeler effective potential Eq.~\eqref{eqRW} are presented in the first four rows denoted by $x_c=1$.
For the case where the truncation point $x_c \ne 1$, one considers the discontinuous effective potential defined in Eq.~\eqref{Veff2}.}
\centering
\begin{tabular}{c c c}
         \hline\hline
 & $\text{Time}$ & $\omega$ \\
\hline
\hline
$\ell=2$, $n=0$, $x_c=1$ ~~~&~~~$1.3630544s$~~~&~~~$0.74734337257090632000 - 0.17792463450502487213i$\\
$\ell=2$, $n=1$, $x_c=1$ ~~~&~~~$1.5696013s$~~~&~~~$0.69342186870496834210 - 0.54783003105244205450i$\\
$\ell=3$, $n=0$, $x_c=1$ ~~~&~~~$1.3574163s$~~~&~~~$1.19888657696247385307 - 0.18540609599648904692i$\\
$\ell=3$, $n=1$, $x_c=1$ ~~~&~~~$1.3579736s$~~~&~~~$1.16528760050917686669 - 0.56259622953244206554i$\\
$\ell=2$, $n=0$, $x_c=\frac78$ ~~~&~~~$2.2256437s$~~~&~~~$0.72118381528323757148 - 0.14526437691857110317i$\\
$\ell=2$, $n=0$, $x_c=\frac34$ ~~~&~~~$2.0099311s$~~~&~~~$0.79427709945222566094 - 0.14845768854491424716i$\\
$\ell=2$, $n=0$, $x_c=\frac58$ ~~~&~~~$2.0235139s$~~~&~~~$0.80929313483339673263 - 0.25418856480114921422i$\\
$\ell=2$, $n=0$, $x_c=\frac12$ ~~~&~~~$2.2058525s$~~~&~~~$0.72595071360605285944 - 0.35517954712959586564i$\\
$\ell=2$, $n=0$, $x_c=\frac{7}{16}$ ~~~&~~~$2.0241544s$~~~&~~~$0.65578027742723877320 - 0.39161380891235662486i$\\
$\ell=2$, $n=0$, $x_c=\frac{3}{8}$ ~~~&~~~$2.0165090s$~~~&~~~$0.57015231756261350254 - 0.41480639631614376557i$\\
$\ell=2$, $n=0$, $x_c=\frac{5}{16}$ ~~~&~~~$2.0236509s$~~~&~~~$0.47128814128866006774 - 0.42258134582724533855i$\\
$\ell=2$, $n=0$, $x_c=\frac{1}{4}$ ~~~&~~~$2.2088341s$~~~&~~~$0.36073050336278809076 - 0.41321763684510711868i$\\
\hline
\hline
\end{tabular}
\end{table*}

\begin{table*}[ht]
\caption{\label{tab5} The same as Tab.~\ref{tab4}, but the calculations are carried out using $N=61$ grid points.}
\centering
\begin{tabular}{c c c}
         \hline\hline
 & $\text{Time}$ & $\omega$ \\
\hline
\hline
$\ell=2$, $n=0$, $x_c=1$            ~~~&~~~$15.2472186s$~~~&~~~$0.74734336883605386867 - 0.17792463137793106952i$\\
$\ell=2$, $n=1$, $x_c=1$            ~~~&~~~$15.2412353s$~~~&~~~$0.69342199377963178797 - 0.54782975059824533074i$\\
$\ell=3$, $n=0$, $x_c=1$            ~~~&~~~$13.6544789s$~~~&~~~$1.19888657687498023688 - 0.18540609588989520797i$\\
$\ell=3$, $n=1$, $x_c=1$            ~~~&~~~$13.6056102s$~~~&~~~$1.16528760606662382182 - 0.56259622687008326346i$\\
$\ell=2$, $n=0$, $x_c=\frac78$      ~~~&~~~$9.6761155s$~~~&~~~$0.65352798558540919416 - 0.10099405275702134695i$\\
$\ell=2$, $n=0$, $x_c=\frac34$      ~~~&~~~$17.9432631s$~~~&~~~$0.68254162149062249289 + 0.09530980537465729105i$\\
$\ell=2$, $n=0$, $x_c=\frac58$      ~~~&~~~$14.7439155s$~~~&~~~$0.96427810473486481862 - 0.10669151996202438263i$\\
$\ell=2$, $n=0$, $x_c=\frac12$      ~~~&~~~$12.7234769s$~~~&~~~$0.88048291499221433394 - 0.11005608759820246501i$\\
$\ell=2$, $n=0$, $x_c=\frac{7}{16}$ ~~~&~~~$10.6927502s$~~~&~~~$0.88405119887826833508 - 0.02887565029098700138i$\\
$\ell=2$, $n=0$, $x_c=\frac{3}{8}$  ~~~&~~~$10.7702415s$~~~&~~~$0.91546337214360609918 - 0.12793971708132637176i$\\
$\ell=2$, $n=0$, $x_c=\frac{5}{16}$ ~~~&~~~$10.7002586s$~~~&~~~$0.84203582212281730098 - 0.14073794945985523660i$\\
$\ell=2$, $n=0$, $x_c=\frac{1}{4}$  ~~~&~~~$12.7259949s$~~~&~~~$0.67600280391770145075 - 0.13222475004761274085i$\\
\hline
\hline
\end{tabular}
\end{table*}

\begin{table*}[ht]
\caption{\label{tab6} Computation time and numerical results of $n$th low-lying QNMs for $s=-2$, $r_h=1$ and for different $\ell$.
The calculations are carried out by the generalized matrix method using $N=101$ grid points at $P=9$th polynomial degree.
The computation time reported in this table does not include that to pre-evaluate the coefficients of Eq.~\eqref{5a}, which is relatively insignificant compared to the original implementation. 
The results for the original Regge-Wheeler effective potential Eq.~\eqref{eqRW} are presented in the first four rows denoted by $x_c=1$.
For the case where the truncation point $x_c \ne 1$, one considers the discontinuous effective potential defined in Eq.~\eqref{Veff2}.}
\centering
\begin{tabular}{c c c}
         \hline\hline
 & $\text{Time}$ & $\omega$ \\
\hline
\hline
$\ell=2$, $n=0$, $x_c=1$ ~~~&~~~$63.4588069s$~~~&~~~$0.74734336886957361594 - 0.17792463137142406912i$\\
$\ell=2$, $n=1$, $x_c=1$ ~~~&~~~$61.3260212s$~~~&~~~$0.69342077647801080535 - 0.54782810465465370116i$\\
$\ell=3$, $n=0$, $x_c=1$ ~~~&~~~$57.8659512s$~~~&~~~$1.19888657687558364419 - 0.18540609588864613087i$\\
$\ell=3$, $n=1$, $x_c=1$ ~~~&~~~$63.2117516s$~~~&~~~$1.16528759479485994054 - 0.56259619577382677886i$\\
$\ell=2$, $n=0$, $x_c=\frac78$ ~~~&~~~$80.7105030s$~~~&~~~$0.78551254551552428928 - 0.13246686031751633238i$\\
$\ell=2$, $n=0$, $x_c=\frac34$ ~~~&~~~$59.9121933s$~~~&~~~$0.79425325467334612895 - 0.14836989910774544451i$\\
$\ell=2$, $n=0$, $x_c=\frac58$ ~~~&~~~$75.2717619s$~~~&~~~$0.80929316918918010881 - 0.25418852912373342463i$\\
$\ell=2$, $n=0$, $x_c=\frac12$ ~~~&~~~$81.9977985s$~~~&~~~$0.72595071362528750441 - 0.35517954712575804010i$\\
$\ell=2$, $n=0$, $x_c=\frac{7}{16}$ ~~~&~~~$77.1996060s$~~~&~~~$0.65578027742882373033 - 0.39161380891226164140i$\\
$\ell=2$, $n=0$, $x_c=\frac{3}{8}$ ~~~&~~~$82.1287758s$~~~&~~~$0.57015231756282194639 - 0.41480639631611043113i$\\
$\ell=2$, $n=0$, $x_c=\frac{5}{16}$ ~~~&~~~$81.7438313s$~~~&~~~$0.47128814128866279507 - 0.42258134582729391225i$\\
$\ell=2$, $n=0$, $x_c=\frac{1}{4}$ ~~~&~~~$87.6446405s$~~~&~~~$0.36073050336276184455 - 0.41321763684512348934i$\\
\hline
\hline
\end{tabular}
\end{table*}

\begin{table*}[ht]
\caption{\label{tab7} The same as Tab.~\ref{tab6}, but the calculations are carried out using $N=101$ grid points at $P=19$th polynomial degree.}
\centering
\begin{tabular}{c c c}
         \hline\hline
 & $\text{Time}$ & $\omega$ \\
\hline
\hline
$\ell=2$, $n=0$, $x_c=1$            ~~~&~~~$62.4549554s$~~~&~~~$0.74734336883608308323 - 0.17792463137787140146i$\\
$\ell=2$, $n=1$, $x_c=1$            ~~~&~~~$67.9073718s$~~~&~~~$0.69342199374632682743 - 0.54782975051101164977i$\\
$\ell=3$, $n=0$, $x_c=1$            ~~~&~~~$62.4415042s$~~~&~~~$1.19888657687498014568 - 0.18540609588989520932i$\\
$\ell=3$, $n=1$, $x_c=1$            ~~~&~~~$63.1596470s$~~~&~~~$1.16528760606661477471 - 0.56259622687004597132i$\\
$\ell=2$, $n=0$, $x_c=\frac78$      ~~~&~~~$84.1997977s$~~~&~~~$0.78469143275473123879 - 0.13326576926007785379i$\\
$\ell=2$, $n=0$, $x_c=\frac34$      ~~~&~~~$64.7049377s$~~~&~~~$0.79425298410996018584 - 0.14836993022114983694i$\\
$\ell=2$, $n=0$, $x_c=\frac58$      ~~~&~~~$74.3326121s$~~~&~~~$0.80929316810172814996 - 0.25418852805030909336i$\\
$\ell=2$, $n=0$, $x_c=\frac12$      ~~~&~~~$86.5767501s$~~~&~~~$0.72595071361390017149 - 0.35517954711882856940i$\\
$\ell=2$, $n=0$, $x_c=\frac{7}{16}$ ~~~&~~~$81.6196972s$~~~&~~~$0.65578027742727741309 - 0.39161380891221109125i$\\
$\ell=2$, $n=0$, $x_c=\frac{3}{8}$  ~~~&~~~$81.3474251s$~~~&~~~$0.57015231756261312025 - 0.41480639631614231542i$\\
$\ell=2$, $n=0$, $x_c=\frac{5}{16}$ ~~~&~~~$87.7176499s$~~~&~~~$0.47128814128866006192 - 0.42258134582724533015i$\\
$\ell=2$, $n=0$, $x_c=\frac{1}{4}$  ~~~&~~~$91.5718613s$~~~&~~~$0.36073050336278809074 - 0.41321763684510711866i$\\
\hline
\hline
\end{tabular}
\end{table*}

\begin{table}[]
\caption{\label{tab8}Numerical results of the fundamental mode $n=0$ for $s=-2$, $r_h=1$, $\ell=2$ and for different grind numbers $N$ and expansion orders $P$.
The calculations are carried out by the original matrix method (MM0)~\cite{agr-qnm-lq-matrix-02}, that adapted for discontinuous effective potential (MM1)~\cite{agr-qnm-lq-matrix-06} and the generalized matrix method (MM2) proposed in this study.
The results for the original Regge-Wheeler effective potential Eq.~\eqref{eqRW} are presented in the rows denoted by $x_c=1$.
For the cases $x_c \ne 1$, one considers the discontinuous effective potential defined in Eq.~\eqref{Veff2}.
}
\begin{ruledtabular}
\renewcommand\arraystretch{2}
\begin{tabular}{cccccc}
Truncation & Method      & $N $	     &	$15$ 	&	$25$    & $37$   \\
\hline
$x_c=1$            & MM0       & $\omega_{\mathrm{Re}}$ & $0.74734350353449625474$   & $0.74734337257090632000$  &  $0.74734336877803351517$  \\
                         &           & $\omega_{\mathrm{Im}}$ & $-0.17792420360892864151i$  & $-0.17792463450502487213i$ & $-0.17792463140472558256i$  \\
\hline
             &           &  $N $     &	$51$	&  $61$	    &  $81$   \\
\hline
                         & MM0       & $\omega_{\mathrm{Re}}$ & $0.74734336883611446812$   & $0.74734336883605386867$  &  $0.74734336883608414160$  \\
                         &           & $\omega_{\mathrm{Im}}$ & $-0.17792463137692161158i$  & $-0.17792463137793106952i$ & $-0.17792463137787143285i$  \\
\hline\hline
             &           & $P/N$	 &	$9/101$	&  $13/101$ &  $19/101$   \\
\hline
	                     & MM2       & $\omega_{\mathrm{Re}}$ & $0.74734336886957361594$   & $0.74734336883605880751$  &  $0.74734336883608308323$  \\
                         &           & $\omega_{\mathrm{Im}}$ & $-0.17792463137142406912i$  & $-0.17792463137804361948i$ & $-0.17792463137787140146i$  \\
\hline
            &            & $P/N$	& $19/201$  & $21/201$  &  $25/201$   \\
\hline
	                     & MM2       & $\omega_{\mathrm{Re}}$ & $0.74734336883608367172$   & $0.74734336883608367162$  &  $0.74734336883608367159$  \\
                         &           & $\omega_{\mathrm{Im}}$ & $-0.17792463137787139683i$  & $-0.17792463137787139654i$ & $-0.17792463137787139656i$  \\
\end{tabular}
\end{ruledtabular}
\end{table}

\begin{table}[]
\caption{\label{tab9}Continuation of Tab.~\ref{tab8}.
}
\begin{ruledtabular}
\renewcommand\arraystretch{2}
\begin{tabular}{cccccc}
Truncation & Method       & $N $	&	$15$	&	$25$	          & $37$ 	   \\
\hline
$x_c=\frac12$            & MM1       & $\omega_{\mathrm{Re}}$ & $0.72595208791085127195$   & $0.72595071360605285944$  &  $0.81561398128303198155$  \\
                         &           & $\omega_{\mathrm{Im}}$ & $-0.35517982360886581440i$  & $-0.35517954712959586564i$ & $-0.44941821366615479582i$  \\
\hline
             &           &  $N $     &	$51$	&  $61$	    &  $81$   \\
\hline
                         & MM1       & $\omega_{\mathrm{Re}}$ & $0.88048291499221433394$   & $1.28820861610820430056$  &  $1.30211969533319582072$  \\
                         &           & $\omega_{\mathrm{Im}}$ & $-0.11005608759820246501i$  & $0.20355222713074846873i$ & $-0.13844243578190514857i$  \\
\hline\hline
             &           & $P/N$	 &	$9/101$	&  $13/101$ &  $19/101$   \\
\hline
	                     & MM2       & $\omega_{\mathrm{Re}}$ & $0.72595071362528750441$   & $0.72595071361389799237$  &  $0.72595071361390017149$  \\
                         &           & $\omega_{\mathrm{Im}}$ & $-0.35517954712575804010i$  & $-0.35517954711883066802i$ & $-0.35517954711882856940i$  \\
\hline
            &            & $P/N$	& $19/201$  & $21/201$  &  $25/201$   \\
\hline
	                     & MM2       & $\omega_{\mathrm{Re}}$ & $0.72595071361390017150$   & $0.72595071361390017150$  &  $0.72595071361390017150$  \\
                         &           & $\omega_{\mathrm{Im}}$ & $-0.35517954711882856946i$  & $-0.35517954711882856946i$ & $-0.35517954711882856946i$  \\
\end{tabular}
\end{ruledtabular}
\end{table}

\begin{table}[]
\caption{\label{tab10}Continuation of Tab.~\ref{tab9}.
}
\begin{ruledtabular}
\renewcommand\arraystretch{2}
\begin{tabular}{cccccc}
Truncation & Method       & $N $	&	$15$	&	$25$	          & $37$ 	   \\
\hline
$x_c=\frac34$            & MM1       & $\omega_{\mathrm{Re}}$ & $0.79476404230901909157$   & $0.79427709945222566094$  &  $0.79644445973558497819$  \\
                         &           & $\omega_{\mathrm{Im}}$ & $-0.14136156307351769649i$  & $-0.14845768854491424716i$ & $-0.14681058722802496426i$  \\
\hline
             &           &  $N $     &	$51$	&  $61$	    &  $81$   \\
\hline
                         & MM1       & $\omega_{\mathrm{Re}}$ & $0.68254162149062249289$   & $0.84050928004322176950$  &  $0.91913982167600729321$  \\
                         &           & $\omega_{\mathrm{Im}}$ & $-0.09530980537465729105i$  & $-0.12508059051311139395i$ & $0.16028437581187569688i$  \\
\hline\hline
             &           & $P/N$	 &	$9/101$	&  $13/101$ &  $19/101$   \\
\hline
	                     & MM2       & $\omega_{\mathrm{Re}}$ & $0.79425325467334612895$   & $0.79425298301079873999$  &  $0.79425298410996018584$  \\
                         &           & $\omega_{\mathrm{Im}}$ & $-0.14836989910774544451i$  & $-0.14836993541502032123i$ & $-0.14836993022114983694i$  \\
\hline
            &            & $P/N$	& $19/201$  & $21/201$  &  $25/201$   \\
\hline
	                     & MM2       & $\omega_{\mathrm{Re}}$ & $0.79425298413719096285$   & $0.79425298413719049345$  &  $0.79425298413719033067$  \\
                         &           & $\omega_{\mathrm{Im}}$ & $-0.14836993025041110071i$  & $-0.14836993025041283636i$ & $-0.14836993025041292994i$  \\
\end{tabular}
\end{ruledtabular}
\end{table}

\begin{table*}[ht]
\caption{\label{tab11} A comparison between the WKB approximation, continued fraction method (CF), the original matrix method (MM0), and the generalized one (MM2) implemented at different orders.
The calculations are carried out for the first two low-lying modes with $\ell=2$, $8$, and $12$, using the original Regge-Wheeler potential with $s=-2$ and $r_h=1$.}
\centering
\begin{tabular}{cccc c}
         \hline\hline
method & order $N$(/$P$) &~~ $\ell$~~ &~~$n$~~~~& $\omega$\\
\hline
\hline
$\text{WKB}$ & $3$ & $2$ & $0$ & $0.7463241297097949 - 0.17843489444859262 i$\\
$\text{WKB}$ & $6$ & $2$ & $0$ & $0.7472387156850152 - 0.17778195900271834 i$\\
$\text{WKB}$ & $9$ & $2$ & $0$ & $0.7476641362282946 - 0.17734120972303288 i$\\
$\text{WKB}$ & $12$ & $2$ & $0$ & $0.17030859629902553 + 79.99304495838905 i$\\
$\text{CF}$ & $100$ & $2$ & $0$ & $0.74734338640252294366 - 0.17792462507976131814 i$\\
$\text{CF}$ & $150$ & $2$ & $0$ & $0.74734336892361384086 - 0.17792463181125186413 i$\\
$\text{MM0}$ & $ 31$ & $2$ & $0$ & $0.74734336885709698481 - 0.17792463088211157522 i$\\
$\text{MM0}$ & $ 51$ & $2$ & $0$ & $0.74734336883611446812 - 0.17792463137692161158 i$\\
$\text{MM2}$ & $ 151/9$ & $2$ & $0$ & $0.74734336883339769781 - 0.17792463138030959831 i$\\
$\text{MM2}$ & $ 151/19$ & $2$ & $0$ & $0.74734336883608366772 - 0.17792463137787138985 i$\\
\hline
$\text{WKB}$ & $3$ & $2$ & $1$ & $0.6920345133033089 - 0.5498307098006824 i$\\
$\text{WKB}$ & $6$ & $2$ & $1$ & $0.6925931628301241 - 0.5469595333990522 i$\\
$\text{WKB}$ & $9$ & $2$ & $1$ & $0.6935283530026902 - 0.5434164498949003 i$\\
$\text{WKB}$ & $12$ & $2$ & $1$ & $0.8103923079669965 + 420.93321813263054 i$\\
$\text{CF}$ & $100$ & $2$ & $1$ & $0.69341608509169484572 - 0.54783348134071378818 i$\\
$\text{CF}$ & $150$ & $2$ & $1$ & $0.69342154249702360068 - 0.54782967603384325873 i$\\
$\text{MM0}$ & $ 31$ & $2$ & $1$ & $0.69342201397727994386 - 0.54782972432278448776 i$\\
$\text{MM0}$ & $ 51$ & $2$ & $1$ & $0.69342199360663158190 - 0.54782975040338458292 i$\\
$\text{MM2}$ & $ 151/9$ & $2$ & $1$ & $0.69342386081139386441 - 0.54783010638232539503 i$\\
$\text{MM2}$ & $ 151/19$ & $2$ & $1$ & $0.69342199376070958125 - 0.54782975057599365801 i$\\
\hline
$\text{WKB}$ & $3$ & $8$ & $0$ & $3.2123645284817237 - 0.1913422499434094 i$\\
$\text{WKB}$ & $6$ & $8$ & $0$ & $3.2123874557032623 - 0.19134141026749504 i$\\
$\text{WKB}$ & $9$ & $8$ & $0$ & $3.212387459385318 - 0.19134139231240976 i$\\
$\text{WKB}$ & $12$ & $8$ & $0$ & $3.212417190849086 - 0.19129970261423618 i$\\
$\text{CF}$ & $100$ & $8$ & $0$ & $3.2123874565454300627 - 0.1913414020537265887 i$\\
$\text{CF}$ & $150$ & $8$ & $0$ & $3.2123874565454300497 - 0.1913414020537265175 i$\\
$\text{MM0}$ & $ 31$ & $8$ & $0$ & $3.2123874565454298520 - 0.1913414020537263707 i$\\
$\text{MM0}$ & $ 51$ & $8$ & $0$ & $3.2123874565454300497 - 0.1913414020537265175 i$\\
$\text{MM2}$ & $ 151/9$ & $8$ & $0$ & $3.2123874565460272697 - 0.1913414020537601529 i$\\
$\text{MM2}$ & $ 151/19$ & $8$ & $0$ & $3.2123874565454300497 - 0.1913414020537265175 i$\\
\hline
$\text{WKB}$ & $3$ & $8$ & $1$ & $3.199584025855904 - 0.5750121921445366 i$\\
$\text{WKB}$ & $6$ & $8$ & $1$ & $3.199622723641189 - 0.575008056108717 i$\\
$\text{WKB}$ & $9$ & $8$ & $1$ & $3.1996227111281206 - 0.5750075867917537 i$\\
$\text{WKB}$ & $12$ & $8$ & $1$ & $3.1976949240618366 - 0.5743890754183395 i$\\
$\text{CF}$ & $100$ & $8$ & $1$ & $3.1996227282712246015 - 0.5750080404058768470 i$\\
$\text{CF}$ & $150$ & $8$ & $1$ & $3.1996227282712124729 - 0.5750080404058748595 i$\\
$\text{MM0}$ & $ 31$ & $8$ & $1$ & $3.1996227282712244491 - 0.5750080404058771754 i$\\
$\text{MM0}$ & $ 51$ & $8$ & $1$ & $3.1996227282712124778 - 0.5750080404058748596 i$\\
$\text{MM2}$ & $ 151/9$ & $8$ & $1$ & $3.1996227296368517211 - 0.5750080462529119829 i$\\
$\text{MM2}$ & $ 151/19$ & $8$ & $1$ & $3.1996227282712124778 - 0.5750080404058748596 i$\\
\hline
\hline
\end{tabular}
\end{table*}

\begin{table*}[ht]
\caption{\label{tab12} Continuation of Tab.~\ref{tab11}.}
\centering
\begin{tabular}{cccc c}
         \hline\hline
method & order $N$(/$P$) &~~ $\ell$~~ &~~$n$~~~~& $\omega$\\
\hline
\hline
$\text{WKB}$ & $3$ & $12$ & $0$ & $4.7710756165976465 - 0.19194224131799553 i$\\
$\text{WKB}$ & $6$ & $12$ & $0$ & $4.771082761522264 - 0.1919420707516933 i$\\
$\text{WKB}$ & $9$ & $12$ & $0$ & $4.771082761365348 - 0.19194207018560772 i$\\
$\text{WKB}$ & $12$ & $12$ & $0$ & $4.77107934506195 - 0.19194184246825335 i$\\
$\text{CF}$ & $100$ & $12$ & $0$ & $4.7710827615789977713 - 0.1919420699026171077 i$\\
$\text{CF}$ & $150$ & $12$ & $0$ & $4.7710827615789977713 - 0.1919420699026171077 i$\\
$\text{MM0}$ & $ 31$ & $12$ & $0$ & $4.7710827615789977651 - 0.1919420699026170852 i$\\
$\text{MM0}$ & $ 51$ & $12$ & $0$ & $4.7710827615789977713 - 0.1919420699026171077 i$\\
$\text{MM2}$ & $ 151/9$ & $12$ & $0$ & $4.7710827615659558654 - 0.1919420698929053876 i$\\
$\text{MM2}$ & $ 151/19$ & $12$ & $0$ & $4.7710827615789977713 - 0.1919420699026171077 i$\\
\hline
$\text{WKB}$ & $3$ & $12$ & $1$ & $4.762459357253952 - 0.5762775161711791 i$\\
$\text{WKB}$ & $6$ & $12$ & $1$ & $4.762471476360502 - 0.5762766633172679 i$\\
$\text{WKB}$ & $9$ & $12$ & $1$ & $4.762471473344963 - 0.5762766667314699 i$\\
$\text{WKB}$ & $12$ & $12$ & $1$ & $4.762370207474441 - 0.5762819478551934 i$\\
$\text{CF}$ & $100$ & $12$ & $1$ & $4.7624714766407203274 - 0.5762766614717375468 i$\\
$\text{CF}$ & $150$ & $12$ & $1$ & $4.7624714766407203282 - 0.5762766614717375504 i$\\
$\text{MM0}$ & $ 31$ & $12$ & $1$ & $4.7624714766407187034 - 0.5762766614717381629 i$\\
$\text{MM0}$ & $ 51$ & $12$ & $1$ & $4.7624714766407203282 - 0.5762766614717375504 i$\\
$\text{MM2}$ & $ 151/9$ & $12$ & $1$ & $4.7624714704986729868 - 0.5762765822999894579 i$\\
$\text{MM2}$ & $ 151/19$ & $12$ & $1$ & $4.7624714766407203282 - 0.5762766614717375504 i$\\
\hline
\hline
\end{tabular}
\end{table*}

\end{document}